\documentclass[11pt]{article}
\usepackage[a4paper,margin=3cm]{geometry}
\usepackage[utf8]{inputenc} 
\usepackage{graphicx}
\graphicspath{{../figures/}}
\usepackage{tabularx}
\usepackage{xcolor}
\definecolor{darkgreen}{rgb}{0,0.5,0}
\usepackage[bookmarks,bookmarksopen,pdfpagemode=UseOutlines,pdfnewwindow=true,
  hyperfootnotes=false,colorlinks,urlcolor=blue,citecolor=darkgreen]{hyperref}
\usepackage{amsmath,amssymb,amsfonts,multicol,multirow,rotating,nicefrac}

\makeatletter
\newcommand{\mythanks}[1]{\hbox{\@textsuperscript{\normalfont#1}}}
\makeatother

\author{       Pablo Jensen\mythanks{1,2,3,4}
\and Jean-Baptiste Rouquier\mythanks{1,2,5}
\and         Yves Croissant\mythanks{1,4}}
\begin{document}
\footnotetext[1]{Universit\'e de Lyon}
\footnotetext[2]{Institut des Syst\`emes Complexes Rh\^one-Alpes (IXXI)}
\footnotetext[3]{Laboratoire de Physique, \'Ecole Normale Sup\'erieure de Lyon and CNRS, 69007 Lyon, FRANCE}
\footnotetext[4]{Laboratoire d'\'Economie des Transports, Universit\'e Lyon 2 and CNRS, 69007 Lyon, FRANCE}
\footnotetext[5]{Laboratoire d'Informatique du parall\`elisme, \'Ecole Normale Sup\'erieure de Lyon and CNRS, 69007 Lyon, FRANCE}

\title{\textbf{Testing bibliometric indicators by their prediction of scientists promotions}}

\maketitle

\begin{abstract}
We have developed a method to obtain robust quantitative bibliometric indicators for several thousand scientists. This allows us to study the dependence of bibliometric indicators (such as number of publications, number of citations, Hirsch index...) on the age, position, etc. of CNRS scientists. Our data suggests that the normalized $h$ index ($h$ divided by the career length) is not constant for scientists with the same productivity but differents ages. 

We also compare the predictions of several bibliometric indicators on the promotions of about 600 CNRS researchers. Contrary to previous publications, our study encompasses most disciplines, and shows that no single indicator is the best predictor for all disciplines. Overall, however, the Hirsch index $h$ provides the least bad correlations, followed by the number of papers published. It is important to realize however that even $h$ is able to recover only half of the actual promotions. The number of citations or the mean number of citations per paper are definitely not good predictors of promotion.
\end{abstract}

\section{Introduction}

A former president of the German Research Foundation declared in 1989: ``When I came to G\"ottingen in 1931, everyone there knew who were the great scientists among the professors [...] and the best young scientists, those with a great future. [...] I still find no fault in this system [of evaluation]. But I know that today it can no longer work effectively... it is an informal system which requires unselfishness and self-criticism on the part of its main participants. This renders it defenseless {\it against suspicion} -- one cannot say: ``I am honest so you must believe me'''' (my emphasis)~\cite{leibnitz}. Prof. Leibnitz nicely summarizes the permanent tension between an expert -- partially subjective -- way of evaluating research and a more objective way, based on (bibliometric) quantitative indicators.

This tension has known renewed interest lately thanks to the introduction of a promising bibliometric indicator, $h$, called the Hirsch index after physicist J. Hirsch who introduced it in 2005~\cite{hirsch}. The h index is defined as the highest number of papers of a scientist that received h or more citations each, while the other papers have not more than h citations each. He suggests that his index reduces several well-known problems of other indices such as the number of articles or the number of citations. Hirsch also introduced the normalized index $h_y$. It represents $h$ divided by the ``scientific age'', i.e. the career length of the scientist, which by convention starts the year of his first publication. This normalization takes into account the fact that $h$ automatically increases with time, and implicitly assumes that $h$ increases linearly in time, an assumption which is not supported by our data (section~\ref{age}).

In this paper, we present a method to extract bibliometric indicators for several thousand scientists. This allows us to study average trends of academic productivity from most scientific domains. Our large dataset also allows us to test empirically two fundamental questions for bibliometry: is $h$ better than the other quantifiers of academic activity in predicting promotions of CNRS researchers to senior positions? Second: if $h$ is the best predictor, is it actually good? By this we mean: what is the proportion of promotions that are predicted by $h$?

Our study significantly improves preceding empirical studies carried out on small samples for technical reasons (difficulty of obtaining large sets of robust bibliometric indicators). Moreover, these studies were generally limited to a subdiscipline: physics~\cite{hirsch}, biomedicine~\cite{bornmann05,bornmann07}, information science~\cite{cronin06}, business~\cite{saad06} and chemistry~\cite{vanraan06}.

On a more general note, our paper wants to contribute empirically to the old discussion of the relevance of bibliometric indicators to account for scientific merits. To be schematic, there are two opposing positions. First, science experts argue that the only way to judge scientists' works and merits, in order to hire or promote them, is through subjective expertise by insiders. Second, some argue that these experts cannot be trusted and promote an ``objective'' science evaluation through the use of well-defined, quantitative operations. Before entering the empirical part of our paper, we think that it important to put this debate into context.

\section{Context : the temptation of mechanical objectivity}

\subsection{In society}

There is a general tendency in modern society to quantify aspects of life, to be able to grasp them more easily. This has been summarized long ago in A. N. Whitehead's famous quote: ``Civilization advances by extending the number of important operations which we can perform without thinking about them. Operations of thought are like cavalry charges in a battle -- they are strictly limited in number, they require fresh horses, and must only be made at decisive moments''\cite{whitehead}. Besides scientists' Top 100~\cite{wos}, examples of this tendency include sports players (baseball, basketball...) or chief executive officers according to their fortune. Summarizing complex systems by homogeneous quantities allows a simpler (but controversial) mathematical treatment of complex questions, as when economists transform pollution or time gains into money. As sociologist Bruno Latour points out: ``The universal yardstick of fortune -- money -- simplifies extremely complex social relations. To account for rank hierarchies between Swann and madame Verdurin, one needs all the subtlety of Marcel Proust. Instead, to rank contemporary millionaires, a Fortune journalist will do''\cite{latour_proust}.

\subsection{Porter's interpretation}

It is tempting to interpret this trend towards quantification as a linear progress towards more powerful and more objective scientific methods. However, this view forgets that quantification is intimately related to expertise, and that expertise is a relation between scientific experts and public officials~\cite{porter}. Therefore, in order to understand quantification, it is
important to look at the social basis of authority of the experts.

In our modern society, expertise based on mere intuition and judgement seems obscure and potentially undemocratic. This is because for non experts, a valid judgement and a subjective bias are difficult to distinguish. Therefore, one solution is to ban subjectivity altogether, even if this leads to throw away (part of) the baby with the bath water: mechanical objectivity, i.e. knowledge based on explicit rules, necessary for quantification, is never fully attainable. As Porter argues, ``The appeal of numbers is especially compelling to bureaucratic
officials who lack the mandate of a popular election. Quantification is a way of
making decisions without seeming to decide. Objectivity lends authority to
officials who have very little of their own''~\cite{porter}.

\subsection{CNRS quantification temptations}
\label{cnrs}

When working on bibliometric indicators, one should keep in mind the diffuse
background of mistrust in peer evaluation of research. For example, Hirsch's
original paper~\cite{hirsch} concludes by: ``I suggest that this index may provide a useful
yardstick to compare different individuals competing for the same resource when
an important evaluation criterion is scientific achievement, {\it in an unbiased way}'' (my emphasis). This resonates with the ambition of science administrators to find a single indicator to select and/or promote scientists, without having to rely on more or less uncontrollable commissions. This tension has been going on for years in France, where peer evaluation commissions defend their expertise against various attempts from CNRS administration to implement ``objective'' promotions based on bibliometric indicators. But this tension is of course more general than its CNRS version, as exemplified by the discussions around the Shanghai ranking of universities (see a short presentation of the obvious shortcomings of such a ranking in~\cite{rankuniv}).

Here are some of the arguments offered by the defenders of the ``expert subjective judgement''. In hard or social sciences, any quantifier supposes a theory that gives meaning to it. And, clearly, citation theory is not one of the strongest! If a significant fraction of citations received by a paper acknowledges its importance for the community, there exist many other well-established reasons to cite a work. For example, Mathew's effect (the more cited get more citations) or the fact that one can cite a paper to criticize it. We refer the reader to~\cite{brooks} for one of the rare empirical studies of citers' motivations and ~\cite{kostoff,leydes,liu} for reviews. Scientists know many ways to improve their citation record, and will develop many more if this indicator becomes crucial for their career. For example, a team can decide to include all its members in all publications (since nor the citation record nor $h$ takes into account the number of authors of a paper) and refer to their own publications extensively. They might even focus their citations on those papers which are a few citations short of counting for their $h$. Therefore, it is unfair to suggest, as Hirsch does~\cite{hirsch}, that only peer committees can be biased. Bibliometric indicators are also biased, although in different and, arguably, more systematic ways.

It is surprising that bibliometric indicators can be uncritically taken as giving the ``true'' value of scientific merits, transforming in ``errors'' any deviation of peer judgment from the $h$ ranking: ``Even though the $h$ indices of approved [...] applicants on average (arithmetic mean and median) are higher than those of rejected applicants (and with this, fundamentally confirm the validity of the funding decisions), the distributions of the $h$ indices show in part overlaps that we categorized as type I error (falsely drawn approval) or type II error (falsely drawn rejection)''~\cite{bornmann07}\footnote{Note that this citation is a clear example of undeserved ``bibliometric'' credit...}.

Needless to say, even if bibliometric indicators could account for academic achievement, they ignore many dimensions of scientists' work which should be taken into account for a proper evaluation : conceptual innovation capacity, risk taking, industrial collaborations, teaching abilities, popularization activities, team management, etc.~\cite{rose,mode2}.

\section{Methods : obtaining reliable bibliometric indicators for several thousand scientists}
\label{summ}

Briefly stated, our method uses the ``Author search'' of the Web of Science (WoS)~\cite{wos} on
the subset of 8750 scientists having filled out the CNRS report the
last three years\footnote{We use data from the annual ``Comptes Rendus Annuels des Chercheurs (CRAC)'' kindly communicated by CNRS Human Resources Department.}. We exclude researchers in Social Sciences (their bibliographic record is not well documented in the WoS) and in High Energy Physics (too few records in the CNRS database), leading to 6900 names. After filtering records suspected to be erroneous, we obtain a database of 3659 scientists with reliable bibliometric indicators, as checked by close inspection of several hundred records. A summary of the subdisciplines encompassed by our database, together with some characteristic average values, is shown in table~\ref{section}. The different positions of CNRS scientists are, by increasing hierarchical importance : ``Charg\'e de Recherche 2$\null^\text{e}$ classe'' (CR2), ``Charg\'e de Recherche 1$\null^\text{re}$ classe'' (CR1), ``Directeur de Recherche 2$\null^\text{e}$ classe'' (DR2), ``Directeur de Recherche 1$\null^\text{re}$ classe'' (DR1) and ``Directeur de Recherche de Classe Exceptionnelle'' (DRCE).

\begin{table}[h]
\begin{center}
  \caption{Overview of the CNRS subdisciplines.}
  \small
  \begin{tabular}{|cp{4.89cm}rp{1.5cm}p{1.1cm}p{0.7cm}p{0.8cm}|}
    \hline
    &Subdiscipline&\raisebox{-0.8\height}{\rotatebox{90}{Section}}
    &Population\newline after\newline filtering&Avg. nb of articles&Avg. $h_y$&Avg. age\\
    \hline
    \multirow{6}{*}{\rotatebox{90}{Physical sciences\hspace{0.5cm}}}
    &Mathematics&1&113&24.79&0.4&42.25\\
    &Physics, theory and method&2&111&50.05&0.77&48.35\\
    &Interactions, particles and strings&3& \multicolumn{4}{c|}{\emph{Not included in our study}}\\
    &Atoms and molecules,\newline lasers and optics&4&150&58.23&0.78&46.24\\
    &Condensed matter:\newline organization and dynamics&5&144&49.71&0.71&46.50\\
    &Condensed matter: structure&6&130&61.99&0.74&47.01\\
    \hline
    \multirow{4}{*}{\rotatebox{90}{Engineering\hspace{0.7cm}}}
    &Information science\newline and technology&7&103&25.85&0.47&41.79\\
    &Micro and nano-technologies, electronics and photonics&8&148&48.95&0.60&44.14\\
    &Materials and structure\newline engineering&9&65&32.82&0.42&46.09\\
    &Fluids and reactants:\newline transport and transfer&10&156&36.01&0.48&46.52\\
    \hline
    \multirow{6}{*}{\rotatebox{90}{Chemistry\hspace{1.35cm}}}
    &Super and macromolecular systems, properties and functions&11&133&52.88&0.73&46.44\\
    &Molecular architecture synthesis&12&122&53.96&0.71&46.86\\
    &Physical chemistry:\newline molecules and environment&13&141&57.92&0.76&46.89\\
    &Coordination chemistry:\newline interfaces and procedures&14&150&59.64&0.80&46.73\\
    &Materials chemistry:\newline nanomaterials and procedures&15&161&64.37&0.67&46.84\\
    &Biochemistry&16&164&56.32&0.73&47.61\\
    \hline
    \multirow{4}{*}{\rotatebox{90}{\parbox{2.1cm}{\centering Earth\newline sciences,\newline astrophysics}}}
    &Solar systems and the universe&17&127&52.94&0.81&47.42\\
    &Earth and earth plants&18&117&40.40&0.67&45.92\\
    &Earth systems: superficial layers&19&80&40.86&0.77&45.89\\
    &Continental surface\newline and interfaces&20&71&37.32&0.69&46.72\\
    \hline
    \multirow{10}{*}{\rotatebox{90}{Life sciences\hspace{2.4cm}}}
    &Molecular basis and structure\newline of life systems&21&153&42.34&0.78&46.64\\
    &Genomic organization,\newline expression and evolution&22&156&40.63&0.79&46.74\\
    &Cellular biology:\newline organization and function&23&133&48.46&0.78&47.67\\
    &Cellular interaction&24&144&46.60&0.83&47.07\\
    &Molecular and integrative\newline physiology&25&140&52.81&0.76&48.09\\
    &Development, evolution,\newline reproduction and aging&26&127&40.64&0.73&47.61\\
    &Behavior, cognition and brain&27&107&37.05&0.67&45.88\\
    &Integrative vegetal biology&28&114&34.78&0.69&47.01\\
    &Biodiversity, evolution\newline and biological adaptation&29&90&47.13&0.81&45.95\\
    &Therapy, pharmacology\newline and bioengineering&30&103&56.89&0.84&45.51\\
    \hline
  \end{tabular}
  \label{section}
\end{center}
\end{table}

\subsection{Detailed description of our procedure}

In the following, we detail our procedure to obtain a large but reliable sample ($\simeq~3\,500$ records) of bibliometric indicators (number of publications, citations and $h$ index). The difficulty lies in the proper identification of the publications of each scientist. Two opposite dangers arise. The first one consists in including extra publications because the request is not precise enough. For example, if only surname and name initials are indicated to WoS, the
obtained list may contain papers from homonyms. The second one consists in
missing some papers. This can happen if scientists change initials from time to
time, or if the surname corresponds to a woman who changed name after marriage.
But this can also happen when one tries to be more precise to correct for the
first danger, by adding other characteristics such as scientific discipline or
French institutions for CNRS scientists. The problem is that both the records
and the ISI classifications are far from ideal: the scientific field can be
confusing for interdisciplinary research, the limitation to French institutions
incorrect for people starting their career in foreign labs, etc.

Basically, our strategy consists in guessing if there are homonyms (see below
how we manage to get a good idea on this). If we think there are no homonyms,
then we count all papers, for any supplementary information (and the resulting
selection) can lead to miss some records. If we guess that there are homonyms,
then we carefully select papers by scientific domain and belonging to French
institutions. After all the bibliometric records have been obtained in this way,
we filter our results to eliminate ``suspect'' records by two criteria: average
number of publications per year and scientist's age at the first publication.

\subsubsection{Evaluate the possibility that there exist homonyms}
\label{homonyms-probability}
\begin{figure*}[hp]
\begin{center}
\includegraphics[angle=-90,width=\textwidth]{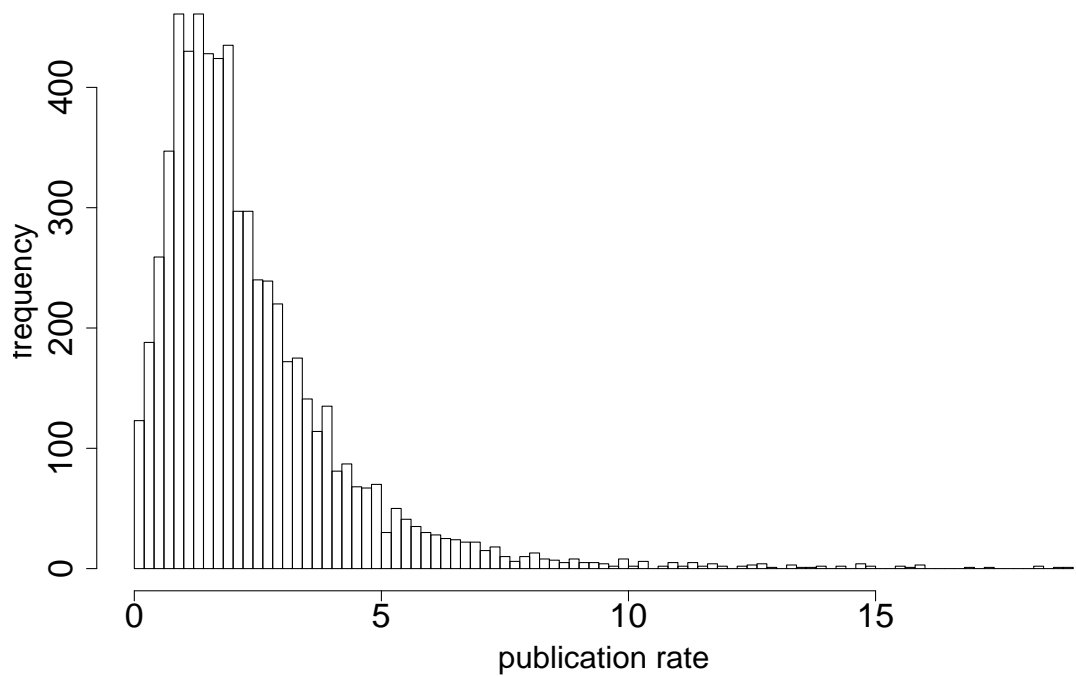}
\caption{Histogram of the scientists publication rate. The rate is calculated as the ratio between the number of publications retained at step (2) and the scientist's career length.}
\label{historate}
\end{center}
\end{figure*}

For this, compute the ratio of the number of papers found for the exact spelling
(for example JENSEN P.) and all the variants proposed by WoS (JENSEN P.*,
meaning P.A. P.B., etc.). If this ratio is large (in our study, larger than .8),
then the studied surname is probably not very common and the author might be the
single scientist publishing. To get a more robust guess, we use the scientist's
age. We look at the total number of papers and compare it to a ``maximum''
normal rate of publishing, taken to be 6 papers a year (see
figure~\ref{historate}). If the publishing rate is smaller than our threshold,
this is a further indication that there is a single scientist behind all the
records. Actually, our strategy can be misleading only when there are only
homonyms with the same initial {\it and} all the homonyms have published very
few papers.

\subsubsection{Obtain the bibliometric records}
\label{obtain-biblio-record}
\paragraph{No homonyms}
If we guess that there is a single scientist behind the publications
obtained for the surname and initials (which happens for about 75\% of
the names), we record the citation analysis corresponding to all
associated papers.

\paragraph{Homonyms}
If we estimate that there are homonyms, we try to eliminate them by using
supplementary data we have. We refine the search by scientific field (``Subject
category'' in WoS terms, but one can select only one) and by selecting only
French institutions\footnote{Unfortunately, WoS allows the selection to be made
  only on the institutions of all coauthors as a whole. So we might retain
  articles of homonyms that have coauthored a paper with a French scientist.}.

\subsubsection{Eliminate suspect records}
\begin{figure*}[hp]
\begin{center}
\includegraphics[angle=-90,width=.9\textwidth]{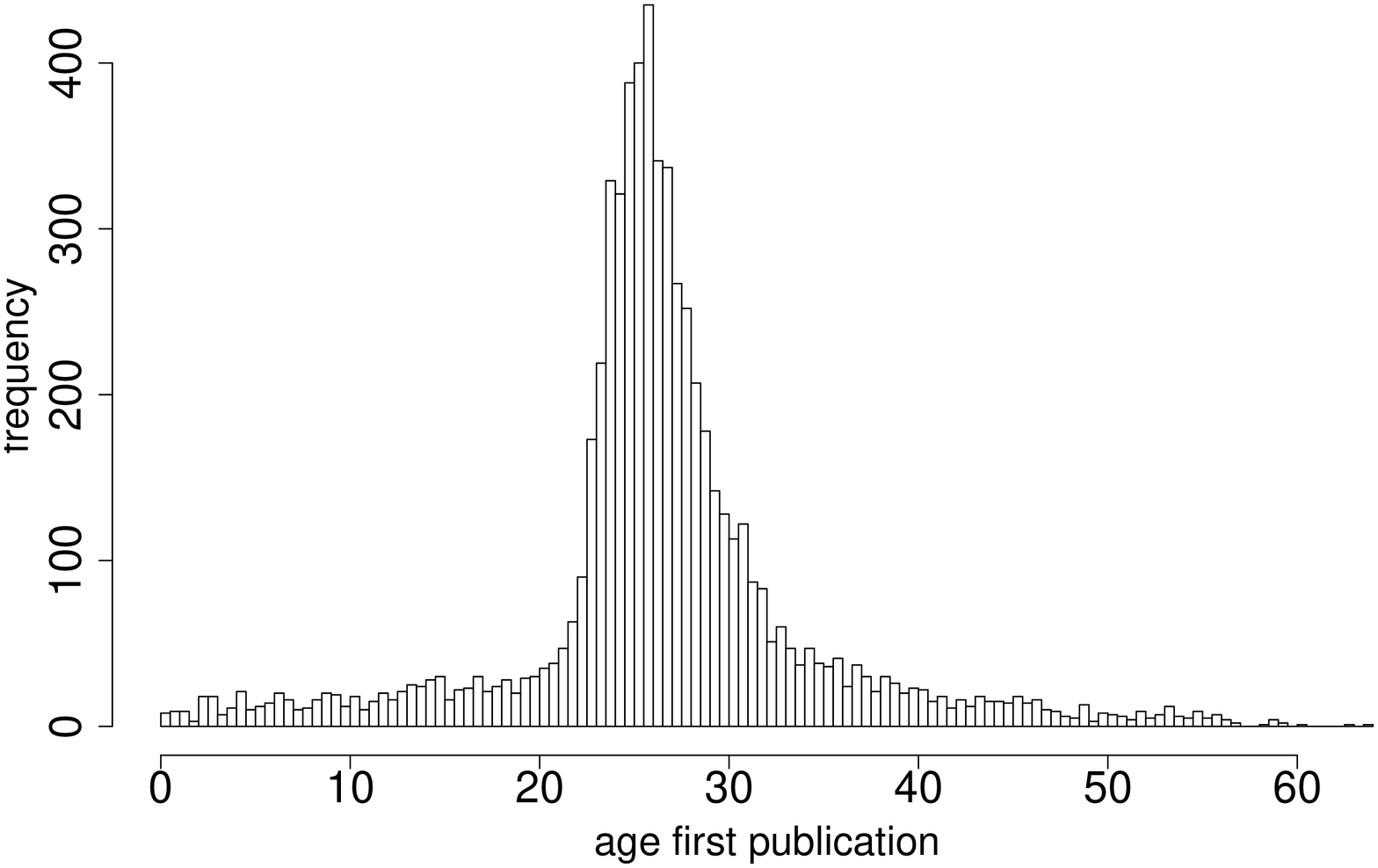}
\caption{Histogram of the scientists age at first publication as calculated from scientist's birth date and publication year of the oldest WoS record retained for that scientist. Clearly, there is a peak at correct ages (i.e. more than 20 and less than 35). Fixing precise thresholds is somewhat arbitrary, but the histogram shows that limits of 20 and 31 years old are not absurd. The results presented here are not qualitatively changed by changing the limits by a few years.}
\label{histoage}
\end{center}
\end{figure*}

Finally, once all the data has been gathered according to the preceding
steps, we eliminate
``suspect'' results by two criteria related to the scientist's age. For a record
to be accepted, the age of the first publication has to be between 21 and 30
years (see figure~\ref{histoage}), and the average number of publications per
year between 0.4 and 6 (see figure~\ref{historate}). After this filtering
process, we end up with 3659 records out of the 6900 initial scientists, i.e. an
acceptance rate of 53\%.

Can we understand why half of the records are lost? First of all, let us detail
how the different filters eliminate records. Deleting scientists who published
their first paper after 30 years old eliminates 1347 ``suspect'' names, which
are probably related to errors or missing papers in the WoS database, to married
women for whom me miss the first papers published under their own surname and to
people who started their career in non French institutions and had homonyms.
Deleting scientists who published their first paper before 21 years old
eliminates 1235 additional ``suspect'' names, which are probably related to
errors in the WoS database, to scientists with older homonyms which we could not
discriminate. Deleting scientists whose record contained less than .4 papers per
year in average leads to the elimination of 121 names. These wrong records can
be explained by the method missing some publications, as in the case ``first
publication after 30 years old''.  Deleting scientists whose record contained
more than 6 papers per year in average leads to the elimination of further 178
names. These wrong records can be explained by the presence of homonyms we could
not discriminate. Finally, to make our database more robust, we decided to
eliminate records suspect of containing homonyms even after selection of
discipline and institution. This is done by eliminating the 359 scientists for
which the number of papers kept after selection is smaller than 20\% of the
total number of papers for the same surname and initials. In those cases, we do
not trust enough our selection criteria to keep such a fragile record.

\subsection{A robust bibliometric database}
In summary, our method leads to a reliable database of around 3500 scientists from all ``hard'' scientific fields. It only discriminates married women having changed surname. It also suffers from the unavoidable wrong WoS records\footnote{For a noticeable fraction of scientists, WoS records start only in the 1990s, even if there exist much older publications, which can be found for example by Google Scholar.}. We stress that the main drawback of the elimination of half the records is the resulting difficulty in obtaining good statistics. But at least we are pretty sure of the robustness of the filtered database. 

Our filtering criteria are based on homonym detection, age of first publication and publication rate. The first criteria correlates only with scientist's surnames, therefore we can expect that it introduces no bias except for married women. Actually, there is a lower woman proportion after filtering: 24.9\% women in the 3659 selection, against 29.6\% in the 6900 database. This is consistent with the preferential elimination of married women who changed surnames and have an incomplete bibliographical record. 

The two other criteria could discriminate scientific disciplines with lower publication rates or underrepresented in WoS. For example, we see in the following table that more scientists from the Engineering Department have been eliminated in the filtering. The mean age is somewhat lower in the filtered database (46.4 years) to be compared to 46.8 in the whole dataset, probably because the records from older scientists have a higher probability of containing errors.

\vspace{.4cm}

\begin{center}
  \noindent
\begin{tabular}{llllp{1.252cm}p{1.5cm}}
\hline
Discipline & Physics &Engineering & Chemistry & Earth Sciences & Life sciences \\
\hline
$\%$ in 6900 database &16.6 & 15.6 & 22.6 &10.2 &34.7 \\
$\%$ after filtering & 18.2 &13.7 & 23.1 & 9.9 &34.8 \\
\hline
\end{tabular}
\end{center}
\vspace{.4cm}

However, overall, the filtered database is very similar to the initial one. For example, the percentage of candidates to senior positions (see below, section~\ref{promotion}) is 16.0\% in the 3659 selection, against 16.4\% in the 6900 database, and the respective promoted percentages are 4.9\% and 5.0\%. The proportions of scientists from each position (see section \ref{summ}) is also similar (Table below): none of the small differences between the filtered and unfiltered values is statistically significant.

\vspace{.4cm}

\begin{center}
  \noindent
\begin{tabular}{llllp{1.252cm}p{1.5cm}}
\hline
Position & CR2 & CR1 & DR2 & DR1 & DRCE \\
\hline
$\%$ in 6900 database &6.0  & 51.8 & 31.9&9.1 & 1.0 \\
$\%$ after filtering & 6.4 & 50.9 & 32.6 & 9.0 &  0.8\\
\hline
\end{tabular}

\end{center}
\vspace{.4cm}

As noted previously, the robustness of our filtered database is validated by the significantly better indicators found for scientists in higher positions. An even stronger test (because the effect is subtler) resides in testing the correlations of the scientist's age at his(her) first publication with several variables : age, position (the reference being DR2, see section \ref{summ}), subdiscipline (Table \ref{section}) and gender. The results of a simple linear regression are shown below :
\vspace{.4cm}

\begin{center}
  \begin{tabular}{llllll}
\hline
variable&coefficient&sd&p-value\\
\hline
(Intercept) &22.6   &0.401936  &$<$ 2e-16  &***\\
age    &0.055    &0.0067  &2.3e-16 &***\\
sex (M)  &0.17    &0.075   &0.023 &*\\
CR2   &0.63    &0.16  &7.8e-05 &***\\
CR1   &0.41    &0.08  &1.9e-07 &***\\
DR1   &-0.53  &0.12  &1.1e-05 &***\\
DRCE  &-0.79   &0.34  &0.022 &*\\
\hline
\end{tabular}
\end{center}
\vspace{.4cm}

We see a progressive decrease of the age of first publication when a scientist has a higher position (all things being equal, for example scientist's age), an effect that is intuitively appealing but certainly small. The fact that we can recover such a subtle effect is a good indication of the robustness of our procedure. We also recover the intuitive effect of scientist's age (older scientists have begun their career later). The gender effect (men publish their first paper 2 months later than women, all other things being equal) is more difficult to interpret, since it mixes many effects : our discrimination (in the filtering procedure) of married women, the unknown effects of marriage and children on scientists' careers, etc. 

We end this section by a brief comment on the influence of gender in our statistics. First, it should be noted that there is a methodological negative bias in our filtering procedure, since married women having changed surname (and not detected by our filters) have a systematically lower publication record (we miss all their publications under her lady's surname). It comes therefore as no surprise that we find lower publication rates or $h_y$ for women. For example, a linear regression on bibliometric indicators (with the following explanatory variables : sex, age, position and subdiscipline) yields an effect of being a woman equal to -5.3 papers or -1 in $h$. The interpretation of this result is difficult, because of our systematic methodological bias. The effects on scientific productivity of marriage or raising children are moreover under debate for both women and men~\cite{sexe,nodiff}. Therefore, we include gender as an explanatory variable in all our regressions to avoid artifacts, but presently, our data cannot shed light on gender effects on scientific productivity.

\section{Trends of academic productivity}
\label{age}

We now take advantage of our large database to study average trends in the academic productivity of CNRS scientists.

\subsection{Age dependence of academic productivity}

First, a word of caution. Our data should be interpreted with care, since they mix two effects. First, the evolution with scientists' ages and, second, a ``generation'' effect. For
example, the generaion aged 50 today started the career in the 80's, when pressure for publication was smaller than today: therefore we could expect lower average publication rates for these scientists.

\begin{figure*}[hp]
\begin{center}
\includegraphics[width=0.7\textwidth]{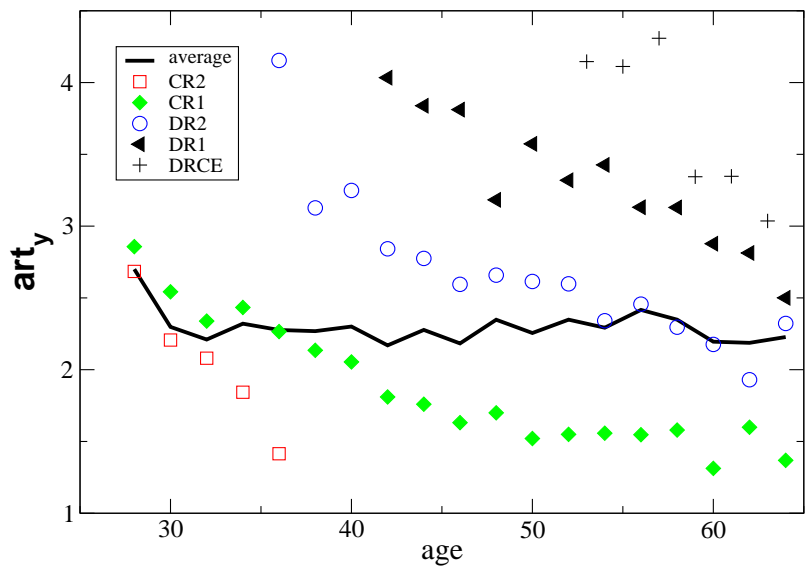}
\caption{Average publication rate as a function of scientist age, on average and for different positions. $art_y$ is obtained by counting the number of papers and dividing by the scientist's career length.}
\label{actyage}
\end{center}
\end{figure*}

\begin{figure*}[hp]
\begin{center}
\includegraphics[angle=-90,width=0.8\textwidth]{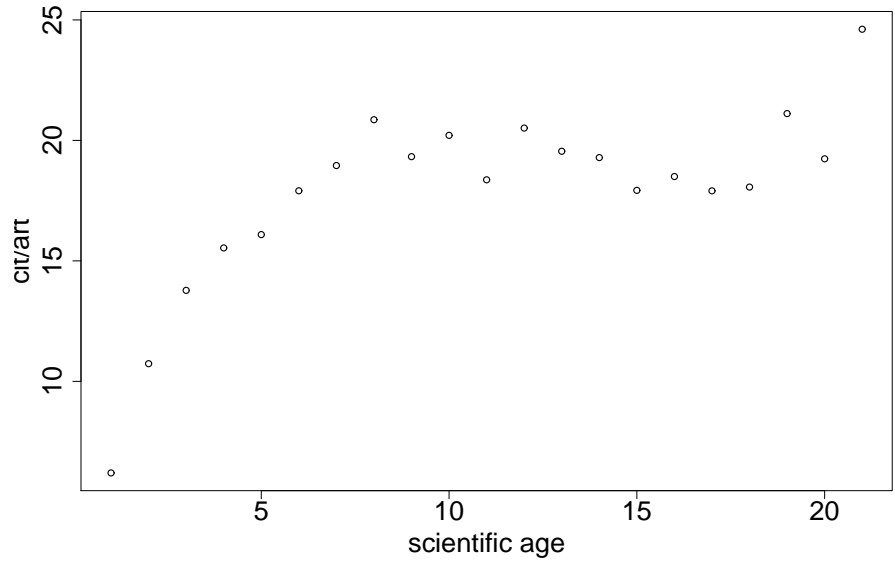}
\caption{Average number of citations per paper as a function of scientific age (i.e. from publication of the first paper). We cumulate all the citations received by the paper since its publication, which disadvantages young scientists with young publications (scientific age less than 7). These data correspond to a mean for all CNRS fields but engineering and mathematics, because the bibliometric characteristics of these fields are relatively different. However, their effect in the means shown here would be negligible.}
\label{citartage}
\end{center}
\end{figure*}

However, the ``generation'' effect does not seem so important when one looks at the evolution of the number of publications per year as a function of age (figure~\ref{actyage}). The figure shows that the average (cumulated) publication rate is remarkably constant with scientist age, and close to 2.2 (see also figure~\ref{historate}). Therefore, our data suggests that scientists produce papers at a steady rate over the course of their careers. Moreover, after a short transient , the mean number of citations per paper is also remarkably constant around 21 citations per paper (figure~\ref{citartage}). This suggests that the scientists produce papers of similar impact over the course of their careers.

\begin{figure*}[hp]
\begin{center}
\includegraphics[width=0.5\textwidth]{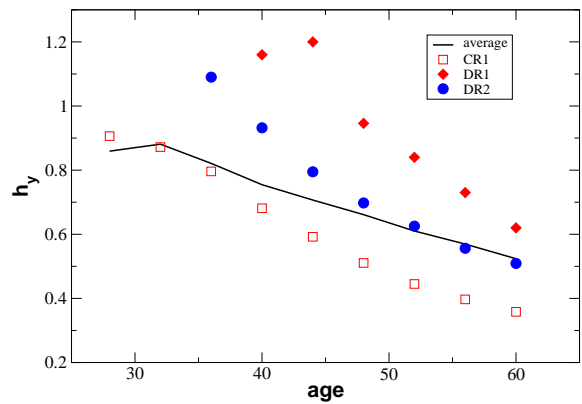}
\caption{Average evolution of $h_y$ as a function of age for different positions.}
\label{hyage}
\end{center}
\end{figure*}

\begin{figure*}[hp]
\begin{center}
\includegraphics[width=0.5\textwidth]{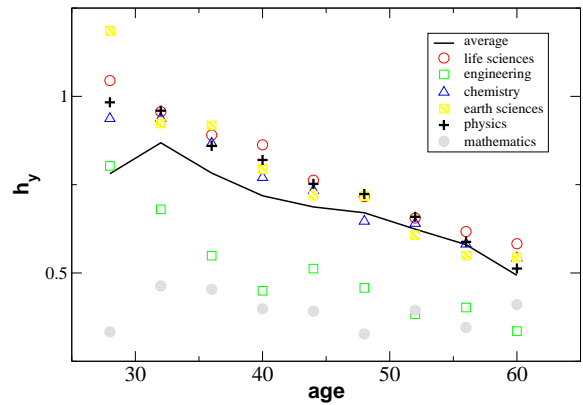}
\caption{Average evolution of $h_y$ as a function of age for different disciplines. Engineering and mathematics have a distinctly lower $h_y$ (partly due to low coverage by WoS of their scientific production), but all the other disciplines have remarkably similar average $h_y$s.}
\label{hyageDS}
\end{center}
\end{figure*}

The time evolution of the average $h_y$ is quite different. Remember that $h_y$ represents $h$ divided by the ``scientific age'', i.e. the career length of the scientist. Figure~\ref{hyage} shows that the average $h_y$ decreases as scientist's age increases, irrespective of scientist's position. This decay is also observed irrespective of scientific field, as shown in figure~\ref{hyageDS}. If one follows the suggestion by Hirsch~\cite{hirsch} that $h_y$ is a good measure to compare scientists of different seniority, it would be tempting to conclude that CNRS scientist's scientific impact decreases with age. For, according to~\cite{hirsch}, a constant  $h_y$ should be observed ``for scientists that produce papers of similar quality (sic) at a steady rate over the course of their careers''. In the following, we show that this interpretation may not be correct, as it assumes a hidden hypothesis which is not born out by bibliometric studies.

To obtain a constant $h_y$, Hirsch~\cite{hirsch} assumes that "the [average] researcher publishes $p$ papers per year and each published paper earns $c$ new citations per year every subsequent year." Then it is easy to show that the combined effect of these two cumulative phenomena (more papers each year, each receiving more citations) leads to a square increase in the total number of citations and a linear increase of $h$. However, the assumption of a constant citation rate {\it unlimited in time} is not supported by bibliometric data~\cite{decay}, as exemplified by the well-known cited and citing half-lifes calculated by the WoS \cite{wos}. Instead, if one assumes a constant publication rate and a constant average impact of these papers (as suggested by our data) but a limited citation lifetime for the average paper, it is easy to show that $h_y$ decreases after a constant transient, basically when $h$ reaches the average number of citations per paper. Therefore, the decrease of $h_y$ observed in our data does not necessarily mean that the impact of CNRS scientists decrease with age. We are currently developing a more sophisticated model based on quantitative citation lifetimes, which could take into account some generation effects to fit quantitatively our data. Therefore, as a first-order approximation, our data suggests that generation effects are negligible and that CNRS scientists productivity is constant in time.

\section{Testing bibliometric indicators for career prediction}
\label{promotion}

There are two issues here: first, comparing the ability of different indicators to predict scientists promotions (i.e. relative performance). Second, comparing the predicted promotions based on the bibliometric indicators to the actual promotions (i.e. absolute predictive performance). In summary, there are two questions: which of the indicators is the best in predicting promotions? Second: is it a good predictor?

\subsection{A brief account of CNRS promotions mechanisms}

Before discussing our results, a brief summary on the CNRS promotion mechanisms may be welcome. The typical CNRS career is as follows. Young scientists enter CNRS around 25-30 years old as ``Charg\'e de Recherche 2$\null^\text{e}$ classe'' (CR2). Then, four years later, they get promoted, almost automatically,  ``Charg\'e de Recherche 1$\null^\text{re}$ classe'' (CR1). A significant fraction end their career in that same position, some 35 years later. Around 35-45  years old (depending on the scientific field), CNRS scientists (in the CR1 positions) start candidating to a senior position : ``Directeur de Recherche 2$\null^\text{e}$ classe'' (DR2). This is the most important career step in CNRS, at least in terms of quantity of scientists involved, and is the main focus of our paper. Each year, less than 10\% of the candidates are promoted. Scientists who have reached the DR2 position can start candidating to ``Directeur de Recherche 1$\null^\text{re}$ classe'' (DR1) positions. A few succeed, at a mean age of 52 years, and a handful of CNRS scientists (less than 1\%) end their career as ``Directeur de Recherche de Classe Exceptionnelle'' (DRCE), the top CNRS position.

\subsection{Bibliometric differences between promoted and non-promoted}

First, let us look at the average differences between promoted and non promoted scientists. Table~\ref{promoave} shows the main differences. We find that all the standard bibliometric indicators strongly correlate with the promotion probability. It has recently been suggested~\cite{iglesias} that the quantity $h/art$ could be a good quantification of the quality of the research.
Basically, this fraction indicates the proportion of ``important'' papers produced by the scientist. However, this indicator is not even significant in predicting the promotions, probably because the number of papers is itself too strong a (positive) indicator of promotion.

\begin{table}[hp]
\begin{center}
\caption{Differences in average values of bibliometric indicators for promoted and non promoted candidates to senior positions. The p-values give the statistical significance of the differences. They are obtained by a standard ``Welch Two Sample t-test''. Standard significance codes for the p-values have been used:  0 ``***'' 0.001 ``**'' 0.01 ``*'' 0.05 and ``.'' for 0.1.}
\medskip
\begin{tabular}{|l|c|c|c|c|}
\hline
Characteristic & promoted & non promoted & p-value\\
\hline
$h$ 	& 15.2 & 12.7 &  1.7 $10^{-7}$ ***\\
$h_y$ 	& 0.85 & 0.73 &  1.7 $10^{-4}$ ***\\
Number of publications	&	49 & 42 & 4.4 $10^{-5}$ *** \\
Number of citations	&		912 & 654 & 3.4 $10^{-5}$ *** \\
Mean citations per paper& 19 & 16 & 0.03  * \\
$h$ / number of papers & .338 & .341 & .79  \\
\hline
Age	&	44.2 & 44.2 & .96 \\
Women \%	&	26 & 21 & .26 \\
\hline
\end{tabular}
\label{promoave}
\end{center}
\end{table}

\subsection{A binomial regression model for the promotion probability}

To test the relevance of the different bibliometric indicators, we analyze the correlation between the promotions of CNRS researchers to senior positions (``Directeur de Recherche 2$\null^\text{e}$ classe'', DR2) and bibliometric quantitative indicators. We add other potentially important variables in our possession, such as subdiscipline (see table~\ref{section}), gender and age. Specifically, we have the list of candidates to senior positions (DR2) over 2004--2006 and the list of promoted scientists. We explain the variable ``promotion'' (1 for the 179 promoted, 0 for the 407 non promoted) for the 586 candidates with a logit model. 

Table~\ref{promoall} confirms the importance of bibliometric indicators: except for the mean number of citations per paper, they are all highly significant. The age has a small influence : for a 45-years old scientist (i.e. the average age for candidates), being a year older increases its promotion probability by only 0.5\%. Note also that, since we control for scientist's age, we expect similar results for both $h$ and $h_y$.

\begin{sidewaystable}[hp]
\centering
\caption{Binomial regressions to explain promotions to senior positions (``Directeur de Recherche 2$\null^\text{e}$ classe'', DR2) on the 586 candidates from all scientific disciplines. The explanatory variables are: sex, age, subdiscipline (table~\ref{section}) and a single bibliometric quantifier ($h$, $h_y$, number of papers (art), number of papers per career length in years ($art_y$), number of citations (cit), average number of citations per paper (citart) or ratio of important papers over all papers (h/art). We also take into account the disciplines (not shown). The columns give the coefficients of the fit for each scientific domain, together with their significance. The last column gives the log likelihood for each of the bibliometric quantifiers. Finally, the last line gives the fit obtained with $h$ as bibliometric indicator for all the 1143 candidates in our database (249 promoted), without filtering. The coefficients are similar, although the bibliometric indicator is less significant, which is consistent with the idea of added noise. Standard significance codes for the p-values have been used:  0 ``***'' 0.001 ``**'' 0.01 ``*'' 0.05 and ``.'' for 0.1.
}
\medskip
\begin{tabular}{|l|l|r|r|r|l@{ }l@{ }l|l|}
\hline
Domain  & intercept & sex (M)    & \multicolumn{1}{c|}{age}& age squared& \multicolumn{3}{c|}{biblio} & logLik\\
\hline
$h$     & -40 (535) & -.22 (.23) & 1.11 (.31) *** &   -.012 (.0034) *** & 0.116   &(.022)    &*** & -328.47\\
$h_y$   & -48 (535) & -.27 (.23) & 1.43 (.32) *** &  -.0147 (.0035) *** & 1.98    &(.38)     &*** &-329.32\\
art     & -41 (535) & -.19 (.23) & 1.17 (.31) *** &   -.013 (.0034) *** & 0.0237  &(.005)    &*** & -332.14\\
$art_y$ & -38 (535) & -.20 (.23) & 1.31 (.30) *** &  -.0138 (.0033) *** & 0.44    &(0.095)   &*** &-336.48 \\
cit     & -38 (535) & -.19 (.23) & 1.06 (.30) *** &  -.0115 (.0033) *** & 0.00058 &(0.00016) &*** &-336.48 \\
citart  & -38 (535) & -.13 (.22) & 1.08 (.30) *** &  -.0116 (.0033) *** & 0.0094  &(.0072)   &    &-342.91 \\
$h$/art & -38 (535) & -.11 (.22) &  1.1 (.31) *** &   -.012 (.0034) *** & -1.52   &(.97)     &    &-342.55 \\
\hline
unfiltered, $h$
        &-16.9 (4.6) ***&-.075 (.16)&.74 (.19) ***& -.0081 (.0021) ***  & 0.045   &(.01)     &*** &-665.1 \\
\hline
\end{tabular}
\label{promoall}
\end{sidewaystable}

\subsection{Comparing the relative relevance of the bibliometric indicators}

\subsubsection{Overall relevance of the bibliometric indicators}

To compare the different bibliometric indicators, the standard tools are the likelihoods of the different fits. These give (table~\ref{promoall}) comparable goodness of fit for $h$ or $h_y$, both significantly better than the fit using the number of articles published or the other indicators. A standard JTest or a CoxTest \cite{jtest} indicate that the models with $h$ or $h_y$ produce a significantly better fit than the model with the number of articles: the p-values are respectively 0.002 (**) and 0.00002 (***).

\subsubsection{Relevance of the bibliometric indicators for different disciplines}

By lack of data, previous studies have focused on a single discipline.  Table~\ref{promofield} examines the relevance of the different indicators for the different fields, the corresponding coefficients being indicated in  table~\ref{promocoeff}. We see that $h$ is the best predictor for all disciplines but Engineering, where the number of papers is more relevant.

\begin{table}[hp]
\begin{center}
\caption{Discipline-specific binomial regressions to explain promotions to senior positions (``Directeur de Recherche 2$\null^\text{e}$ classe''. The explanatory variables are: sex, age, subdiscipline (table~\ref{section} and a single bibliometric quantifier (h, $h_y$, number of citations (cit), number of papers (art) or the average number of citations per paper (citart). The figures in the bibliometric columns refer to the log likelihood of the corresponding regression. To simplify the presentation, we don't show their coefficients, since they are always significant.}
\begin{tabular}{|p{2.55cm}|c|c|c|c|c|c|c|c|}
\hline
Domain & $h$ & $h_y$ & art & cit & citart & candidates & promoted\\
\hline
Physical sciences  & -46.4 & -45.4 & -52.0 & -48.3 & -51.7 & 114 & 30\\
Life sciences & -90.2 & -91.7 & -94.2 & -93.0 & -96.1 & 165 & 58\\
Engineering & -40.5 & -41.2 & -38.9 & -41.6 & -41.3 & 85 & 29\\
Chemistry & -78.7 & -78.8 & -79.1 & -80.8 & -81.5 & 150 & 44\\
Earth Sciences, Astrophysics & -32.9 & -33.2 & -33.3 & -34.4 & -34.6 & 72 & 18\\
\hline
\end{tabular}
\label{promofield}
\end{center}
\end{table}

\begin{table}[p]
\begin{center}
\caption{Binomial regressions to explain promotions to senior positions. The explanatory variables are: sex, age, discipline, dissemination activities (taken as a binary variable: the coefficient refers to ``active'') and subdisciplines (see table~\ref{section}, not shown). We show the coefficients for the bibliometric indicator which gives the best fit (as defined by the likelihood, see table \protect\ref{promofield}). Standard significance codes for the p-values have been used:  0 ``***'' 0.001 ``**'' 0.01 ``*'' 0.05 and ``.'' for 0.1.}
\begin{tabular}{|p{2.55cm}|r@{ }l|r@{ }l|r@{ }l|r@{ }l|}
\hline
Domain & \multicolumn{2}{c|}{sex (M)} & \multicolumn{2}{c|}{age} & \multicolumn{2}{c|}{age  squared} & \multicolumn{2}{c|}{biblio}\\
\hline
Physical sciences            & -1.3 (.7)& .  & 3.3 (1.2)& ** & -.03  (.01) &** & 5.1 (1.3) &*** \\
Life sciences                & .2 (.4) &.    & .46 (.76)&    & -.006 (.008)&   & .15 (.045)& ** \\
Engineering                  & .7 (1)&       & 3.2 (1.4)& *  & -.036 (.016)& * & .038 (.018)& * \\
Chemistry                    &  -.7 (.45)& . & 1.2 (.56)& *  & -.011 (.006)& . & 1.8 (.8)   & * \\
Earth Sciences, Astrophysics & .5 (.8)&      &  1.2 (.8)&    & -.018 (.009)&   & .12 (.07)  &. \\
\hline
\end{tabular}
\label{promocoeff}
\end{center}
\end{table}

\subsection{Testing the absolute predictive power of the best bibliometric indicator}

We have shown that, overall, $h$ is the best bibliometric indicator to account for CNRS promotions. We now want to calculate how good it is, i.e. the absolute performance in accounting for the promotions.

There are many ways to do this. First, one can rank candidates by their bibliometric indicators, promote the top of the list and compare to the actual promoted list. This has to be done by subdiscipline (table~\ref{section}) since promotions are decided at this scale. We find that $h$ ranking leads to 48\% of ``correct'' promoted scientists, while ranking by number of citations gives 46\% and ranking by number of published papers only 42\%. These figures should be compared to a ``random'' ranking which would achieve 30\% of ``correct'' promotions (i.e. the proportion of promoted scientists in a random sample).

Alternatively, one can calculate the average promotion probabilities for promoted and non-promoted: for the $h$ binomial model (table~\ref{promoall}), these are 0.396 and 0.266, while for the number of papers model, we find 0.386 and 0.270. In both cases, promoted have a significant higher probability according to the model, and differences are clearer with $h$, which is consistent with its higher likelihood. Finally, one could estimate the proportion of correct promotion predictions by the binomial model (table~\ref{predic}) as compared to a purely random model: we improve the proportion of correct predictions from 57.6\% to 71\%. Note however that half of the actually promoted candidates are not promoted by the binomial model.

\begin{table}[hp]
\begin{center}
\caption{Number of correct predictions to senior positions promotions. The first two lines indicate the predictions of the binomial model (table~\ref{promoall}), taking as threshold for effective promotion the probability value $.3735$, determined to recover the true proportion of promoted scientists (0.305). The random model is obtained by choosing randomly between promotion (probability $.305$, to recover the same proportion of promoted scientists) and non promotion (probability $1-.305$). The proportion of correct promotions is found by summing the diagonal terms and dividing by the total number of cases.}
\medskip
\begin{tabular}{|rc@{}|}
\hline
& \hspace{2.8cm}reality\\
\multirow{3}{*}{binomial model}
 &
  \multirow{2}{*}{
    \begin{tabular}{|rcc|c}
      \hline
      & non promoted & promoted & total \\
      non promoted 	& 322 & 85  & 407 \\
      promoted      	& 85  & 94  & 179 \\
      \hline
      total          & 407 & 179 & 586 \\
      non promoted 	& 283 & 124 & 407 \\
      promoted       & 124 & 55  & 179 \\
      \hline
      total          & 407 & 179 & 586 \\
    \end{tabular}
  }\\
&\\ &\\ &\\
\hline
\multirow{3}{*}{random model} & \\
&\\ &\\
\hline
\end{tabular}
\label{predic}
\end{center}
\end{table}

As a conclusion, bibliometric indicators do much better than randomness to predict promotions but there remains significant differences. For example, a ``mechanical objectivity'' procedure, which ranks candidates by their $h$ would disagree with actual promotions for half of the promoted people, a very significant difference.

\subsection{Test on DR2 to DR1 promotion}

We end by studying the promotion to the highest CNRS positions, ``Directeur de Recherche 1$\null^\text{re}$ classe'' (DR1). We have data on 67 scientists promoted over 376 candidates. The mean age of candidates is 53 years, with a mean number of publications of 74.6, and a mean $h$ of 18.6 (giving a mean $h_y = .7$). The promoted scientists are 52 years old, with a mean number of publications of 88.6, and a mean $h$ of 22 (giving a mean $h_y = .83$). Table~\ref{dr1} shows the log-likelihood of the binomial fits as a function of the selected bibliometric indicator. Note that the scarcity of the data does not allow a meaningful fit by discipline, as performed in the DR2 case (table~\ref{promofield}). Our data shows that the most relevant predictor of promotion is the number of papers published by the candidate.

\begin{table}[p]
\begin{center}
\caption{Binomial regressions to explain promotions to the highest senior positions (Directeur de Recherche 1$\null^\text{re}$ classe, DR1). The explanatory variables are: sex, age and subdisciplines (see table~\ref{section}, not shown). We show the log-likelihood for the different bibliometric indicators.}
\medskip
\begin{tabular}{|l|c|c|c|c|c|}
\hline
 $h$ & $h_y$ & art & $art_y$ & cit & citart \\
\hline
-147.4 & -148.6 & -145.0 & -148.0 &-152.5 & -155.3 \\
\hline
\end{tabular}
\label{dr1}
\end{center}
\end{table}

J or Cox tests \cite{jtest} give a high significance to the log-likelihood difference between number of articles and $h$ (p-values of 0.01688 * and 0.000665 *** respectively).

\begin{table}[p]
\begin{center}
\caption{Binomial regressions to explain promotions to the highest senior positions (Directeur de Recherche 1$\null^\text{re}$ classe, DR1). The explanatory variables are: sex, age and subdisciplines (see table~\ref{section}, not shown). We show the coefficients for the number of articles, which is the best bibliometric indicator as defined by the log-likelihood. Standard significance codes for the p-values have been used:  0 ``***'' 0.001 ``**'' 0.01 ``*'' 0.05 and ``.'' for 0.1.}
\medskip
\begin{tabular}{|l|c|c|c|c|c|}
\hline
intercept & art & sex (M) & age & age squared \\
\hline
-45 (17) ** & 0.025 (0.0057) *** & -.67 (.42) & 1.66 (.65) * & -.016 (.006) ** \\
\hline
\end{tabular}
\label{dr1fit}
\end{center}
\end{table}

\section{Discussion, Conclusions}

Thanks to a new filtering method, we have obtained a large database of scientists' bibliometric records. We have been able to determine the average trends of bibliometric indicators as a function of scientists' ages, positions and disciplines. Our main result is that generation effects seem negligible, since the mean publication rate or the mean number of citations per paper do not vary significantly with scientists' ages. However, the average normalized Hirsch index $h_y$ does decrease with age, a trend that seems related to the very definition of $h_y$ and not to a decrease of scientific activity.

Our large database has also allowed us to study the relative relevance of bibliometric indicators to predict promotions to senior positions. We find that, overall, $h$ shows the best performance, while the number of papers is second and accounts better for promotions of scientists from the engineering department and for promotions to the highest CNRS positions (``Directeur de Recherche 1$\null^\text{re}$ classe''). Incidentally, the good prediction of promotions by the bibliometric indicators confirms that our large database of bibliometric records is robust. 

To conclude, let us come back to our inital controversy on ''mechanical objectivity'', i.e. the idea of deciding promotions automatically, on the basis of bibliometric indicators. Our study shows that the consequences would be dramatic, changing roughly half the promotions every year. The same order of magnitude for the difference between $h$ and peer committees' rankings can be guessed from Ref.~\cite{bornmann07}. Further studies are needed to understand the differences of the two rankings, which may be surprising since $h$ is already used to evaluate scientists in many CNRS subdisciplines. One could argue that promotion is also determined by scientists' activities which are not taken into account in the bibliometric indicators (see above, \ref{cnrs}). However, in a forthcoming paper~\cite{dissem}, we show that dissemination activities (industrial collaborations, popularization and teaching) are practically irrelevant for peer committees' decisions about promotions. The differences could then be explained by the consideration of additional activities (team management, risk taking \ldots), by unsatisfactory peer evaluation (preferential promotion of friends, visible colleagues\ldots) or by the opposite : human expertise capable to judge whether automatic measures are really meaningful. Automatic ranking appears as a clear political choice, selecting some of the scientists' activities and distrusting peer committes.

\vspace{1cm}

We acknowledge illuminating discussions with Y. Langevin and S. Bauin from CNRS, J. LeMarec from Ecole Normale Supérieure-Lettres et Sciences Humaines and P. Kreimer from Universidad de Quilmes (Argentina). The data have been kindly provided by CNRS Human Resources administration: it is a pleasure to thank Y. Demir, A. Pes and F. Godefroy for their precious help.

\end{document}